# Molecular understanding of charge storage and charging dynamics in supercapacitors with MOF electrodes and ionic liquid electrolytes

Sheng Bi[1,2], Ming Chen[1,3], Runxi Wang[1], Jiamao Feng[1], Mircea Dincă[4], Alexei A. Kornyshev[2*], and Guang Feng[1,*]

We present a 'computational microscopy' analysis (targeted molecular dynamics simulations) of the structure and performance of conductive metal organic framework (MOF) electrodes in supercapacitors with room temperature ionic liquids. The molecular modeling predicts the characteristic shapes of the potential dependence of electrode capacitance, relying on the structure of MOF electrodes and particularly how ions transport and reside in MOFs under polarization. Transmission line model was adopted to characterize the charging dynamics process and build up a bridge to evaluate the capacitive performance of practical supercapacitor devices at macroscale from the simulation-obtained data at nanoscale. Such nanoscale-to-macroscale analysis demonstrates the potential of MOF supercapacitors for achieving unprecedentedly high volumetric energy and power densities. The investigation gives molecular insights into the preferred structures of MOF for achieving these results, which could provide a blueprint for future experimental characterization of these new systems.

[1]State Key Laboratory of Coal Combustion, School of Energy and Power Engineering, Huazhong University of Science and Technology (HUST), Wuhan 430074, China.
[2]Department of Chemistry, Faculty of Natural Sciences, Imperial College London, Molecular Sciences Research Hub, White City Campus, W12 0BZ, London, United Kingdom.
[3]Shenzhen Research Institute of HUST, Shenzhen, 518057, China.
[4]Department of Chemistry, Massachusetts Institute of Technology, Cambridge, MA, 02139, United States.
*e-mail: a.kornyshev@imperial.ac.uk; gfeng@hust.edu.cn

Enhancing capacitive performance of modern **e**lectrical **d**ouble **l**ayer **c**apacitors (EDLCs), also called supercapacitors, crucially relies on the development and application of porous electrode materials.[1, 2] Owing to their tunable porous structures and synthetic advantages,[3, 4, 5] conductive **m**etal **o**rganic **f**ramework (MOF) materials have a great potential for such electrodes.[5, 6, 7, 8] Their scaffold-shaped volume-filling structure could bring a large specific surface area per mass or volume, with a custom-designed pore space.[3, 6] This helps to maximize the capacitance and ultimately the energy density, but may also promote ion transport during charging/discharging, thereby increasing power density. Indeed, graphene-doped MOFs were found to give high capacitance, due to their high porosity and openness of their structure.[9] Furthermore, a highly conductive MOF $Ni_3$(2,3,6,7,10,11-hexaiminotriphenylene)$_2$ ($Ni_3(HITP)_2$) has shown very high areal capacitance and low cell resistance, superior to most carbon-based materials, when used as the sole electrode material for EDLC with an organic electrolyte.[10]

Electrolyte is equally important to the EDLC performance. **R**oom **t**emperature **i**onic **l**iquids (RTILs) emerged as notable candidates for electrolytes because of their excellent thermal stability, nonvolatility, broad working temperature range, and wide electrochemical window;[11, 12, 13] the latter could potentially facilitate boosting the energy density of EDLCs.[13, 14] To build high-performance EDLCs, many studies focused on understanding the energy storage mechanism of porous electrodes with RTILs, via *in situ* experiments and molecular simulations.[11, 15, 16, 17, 18] Traditional porous electrodes, such as activated carbons, have wide pore size distributions,[1, 2, 3] making molecular understanding through simulations difficult, whereas MOFs present monodisperse pores of controllable dimensions,[3, 4, 5, 7] making them near-ideal systems for computational modeling. Nevertheless, because there could be millions of MOF-RTIL combinations, it is important to first unravel generic mechanisms of charge storage and charging dynamics related to their structure, in particular, considering that to date there is no work on supercapacitors coupled with conductive MOFs as sole electrode materials and RTILs as electrolytes.

In this work, we focus on this task, using 'computational microscopy': molecular dynamics (MD) simulations based on atomistic models of MOFs and coarse-grained models of RTILs. As shown in Fig. 1a, our MD system consists of two identical MOF electrodes immersed in a RTIL. We consider three types of electrodes based on densely stacked 2D-conductive MOF sheets with different-sized quasi-1D pores (Fig. 1b-c), and 1-ethyl-3-methylimidazolium tetrafluoroborate ([EMIM][BF$_4$]) as the electrolyte (Fig. 1d). In MD simulations, we control the voltage between the cathode and anode; for each electrode, we calculate the electrode potential as the potential difference between the electrode and the electrolyte reservoir, calibrating it relative to the **p**otentials of **z**ero **c**harge (PZC) of the electrode. For simulation details see Methods. After equilibrating the system at PZC (i.e., 0 V), we apply jump-wise voltages between two electrodes, and then monitor the follow-up charging dynamics, as well as charge and ion distributions after reaching equilibrium, investigating the effect of the value of applied voltage. We explore the structure of ionic distributions in electrically polarized, nanoscale pores and establish the options for energy storage and power delivery that these structures could proffer.

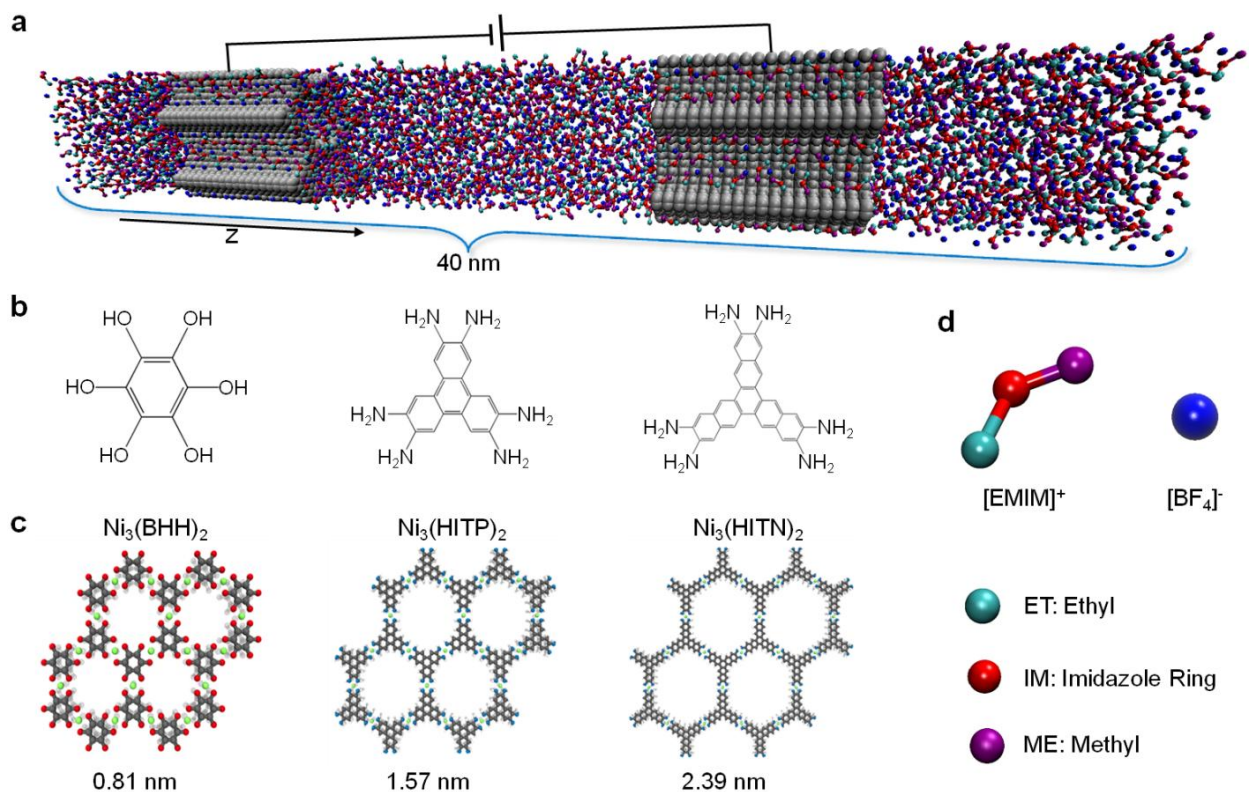

**Fig. 1. Schematics of MD simulation of MOF-based supercapacitors. a,** A snapshot of the simulation system containing two identical MOF electrodes connected with RTIL reservoirs (periodically repeated in all three dimensions). Each electrode has stacks of eighteen 2D MOF sheets. **b,** Molecular structures of the linkers used for three studied MOFs. **c,** 2D honeycomb structures of MOF sheets studied. The numbers at the bottom indicate characteristic in-plane sizes of quasi-1D pores formed by the stacked MOF sheets (i.e., effective pore diameters). **d,** Coarse-grained model of RTIL [EMIM][BF$_4$]. Details for all three studied MOFs and simulation setup can be found in Supplementary Part 1 and Supplementary Tables 1-3.

**Equilibrium charge and ion distributions inside MOFs**

We now present the simulation results for charge and ion distributions in neutral and polarized MOF pores, with first calculating PZC values as 0.074, 0.035 and 0.042 V for studied MOF electrodes with pore diameters of 0.81, 1.57 and 2.39 nm, respectively. The small PZC values suggest that there is no noticeable preferential adsorption of cations or anions of [EMIM][BF$_4$] into the pores of the considered MOFs.[19]

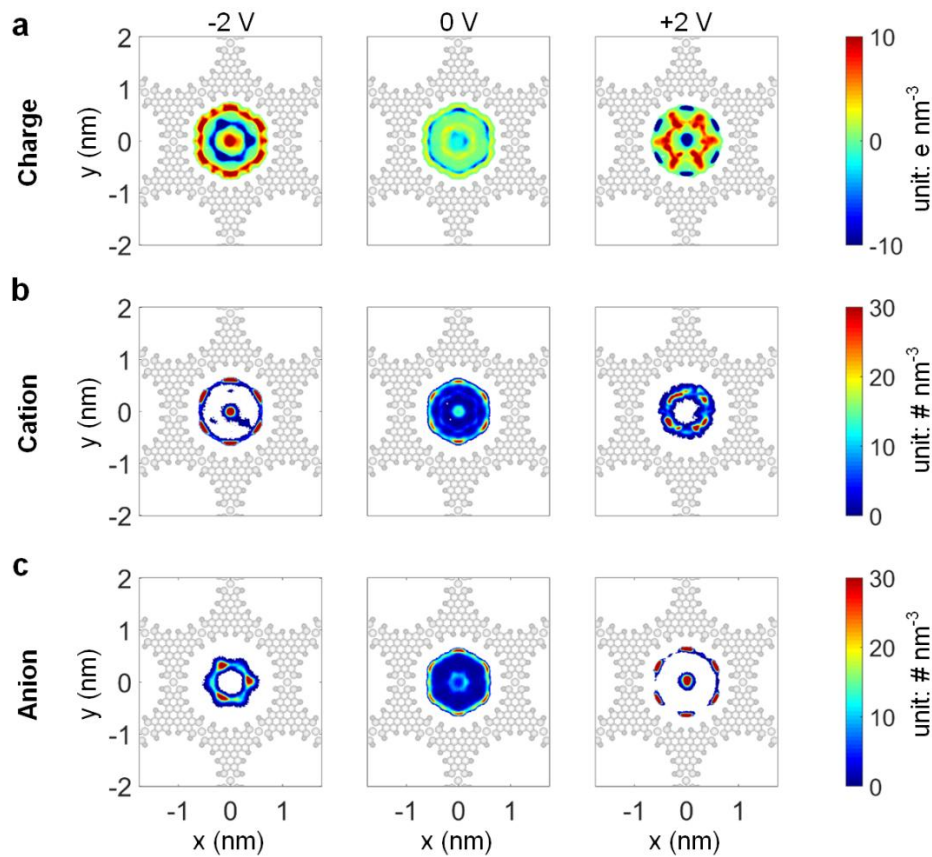

**Fig. 2. In-plane, 2D maps of charge and ion distributions of [EMIM][BF$_4$] inside a pore of a studied MOF.** Each map is based on simulation data averaged along the pore axis (pore diameter is 1.57 nm). Columns correspond to three (indicated) electrode potentials (0 V means PZC, see the main text). **a**, 2D charge distributions. **b-c**, 2D number density distributions of cations (**b**) and anions (**c**). For better visibility, results for neighboring pores are not displayed, and in the central pore the areas where no ions could access are shown in white.

Figure 2 depicts the in-plane charge and ion distributions in a 1.57 nm pore at different electrode potentials. Inside quasi-1D pores of a polarized MOF electrode, radial ion distributions are found to be more heterogeneous than those at PZC, with counter-charge positioned closer to the pore surface and at pore center-line (Fig. 2a). As shown in Fig. 2b-c, at PZC, cations and anions in the pore both form an ion layer adsorbed on the pore surface, showing a hexagonal pattern in the planar-cross-section, with a wire of ions along the pore axis. Under electrode polarization, the *counter-ions* (i.e., cations at negative and anions at positive polarization) exhibit a packing of the same kind but with more distinct separation between a surface-adsorbed layer and the center-line. The *co-ions* (ions of opposite charge to counter-ions) settle between these two regions. For the smaller pore MOF (0.81 nm pore diameter), only one layer of ions can get accommodated inside the pore, regardless of electrode polarization (cf. Supplementary Figure 3). For the larger pore (2.39 nm), at PZC, two mixed cation-anion layers dwell inside the pore, one of them contacting pore wall; with electrode polarization they differentiate into more pronounced two counter-ion layers separated by a co-ion layer, with a co-ion wire at the axis (cf. Supplementary Figure 4).

The interlaced distributions of cations and anions along the radial direction of the polarized pore comply well with the ion layering in RTIL-based EDLs at electrode surfaces revealed by previous experiments and simulations,[11, 15, 16, 17] while there is little cation-anion layering shown at PZC (middle column of Fig. 2). To delve into this difference, the ion density was analyzed along the pore axis. In Fig. 3a we see 'wave-like' axial ion distributions in the pores of the non-polarized electrode, which become even more distinct with electrode polarization. Further looking into ions lodging inside pores, we divided the pore space into central and surface regions demarcated by a critical radial distance, $R_c = 0.25$ nm (Fig. 3b), which are illustrated by simulation snapshots (Fig. 3c). Therefore, the cation orientations, characterized by angular distributions in Fig. 3d, reveal how ions could be accommodated inside the MOF pore. As the electrode gets more negatively charged, cations in the pore surface region prefer to be aligned along the pore axis, while those in the central region distribute more randomly, which could be ascribed to the adsorbed ion layer screening out the electrode surface charge. With more positive polarization, the cations move closer to the center-axis and leave their ethyl groups pointing to the surface-adsorbed anions. Similar trends could be observed for smaller and larger pores (Supplementary Figures 5 and 6), except that ions in the central region of larger pore orientate more randomly. The rich micro-structures, delineated by the in-plane and axial ion distributions, help to understand where and why ions should 'comfortably' reside in the porous electrode.[11, 15, 20]

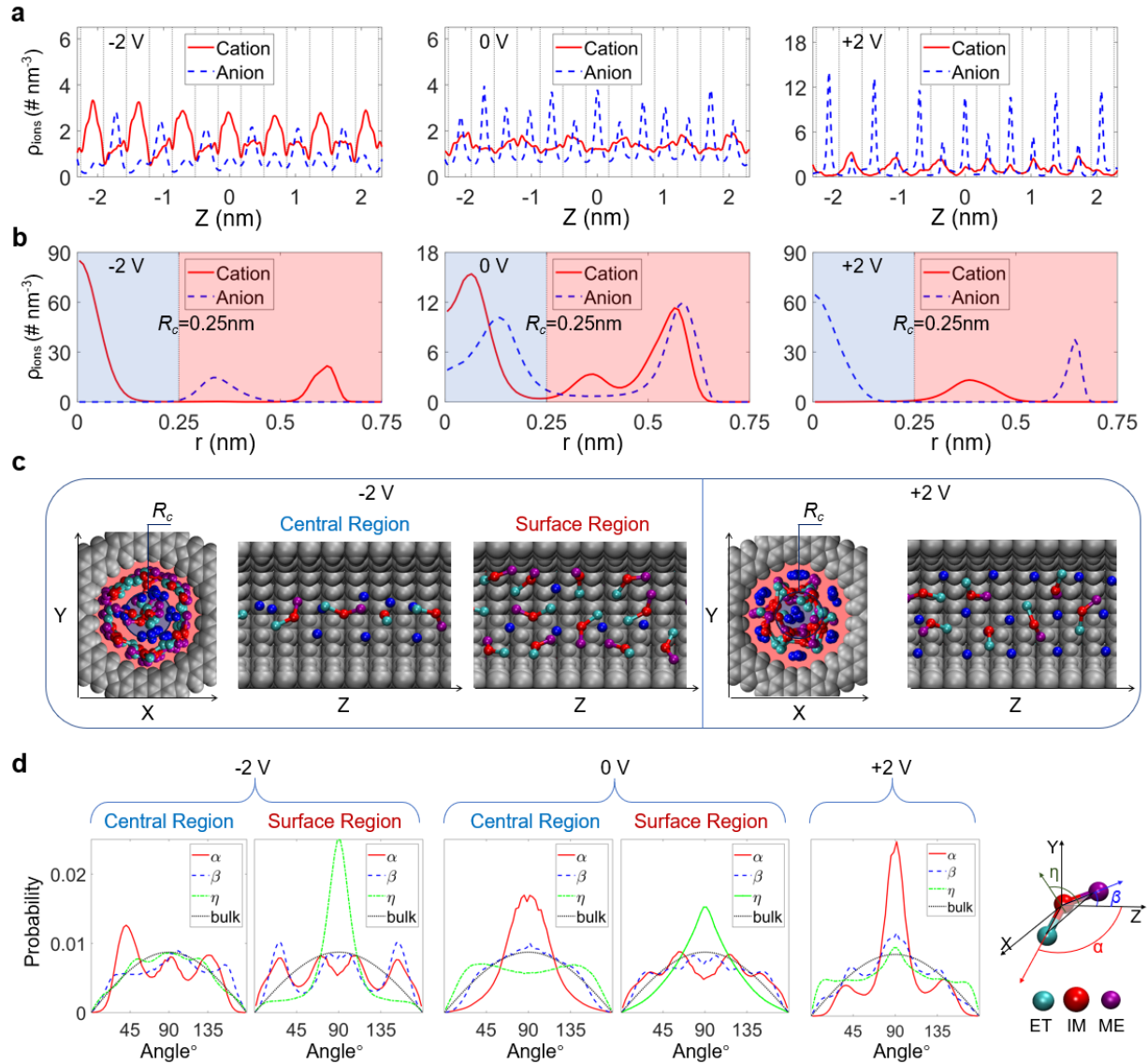

**Fig. 3. In-pore ion density and orientation distributions. a**, Axial distributions of the ion number. **b**, Radial ion distributions. The light blue and red shaded areas, separated by a critical value of radial distance ($R_c$), represent the central and surface regions of the pore space. **c**, Typical snapshots of ions inside the negatively and positively charged MOF pores (colours of ions same as in Fig. 1d). The pore side view of snapshots in negative electrode shows ions, separately, in the central and surface regions defined in **b**. **d**, Angular distribution of cations located in central and surface regions of the MOF pore.

## Capacitance and energy density of MOF-based supercapacitors

Charge storage in supercapacitors is characterized by capacitance and energy density as a function of voltage. Differential capacitance of an individual electrode is defined as the derivative of the electrode's charge with respect to its potential. The charge on the MOF electrode, determined by an excess or depletion of electrons in it, is equal with an opposite sign to the net ionic charge accumulated inside the pores and in the double layer at the electrode's outer surface. For a highly porous electrode, the area of the latter is negligible, and we will not consider that contribution.

In the metallic electrode, before the onset of electrochemical reactions, there are no limitations on its ability to accommodate electrons; the same refers to charging electrodes positively. The capacitance of such electrode is determined by the ability to accumulate the ionic charge from electrolyte. In electrodes built from low dimensional materials, such as graphene and its derivatives, accommodation of electrons has its own laws that give rise to the **q**uantum **c**apacitance (QC) contribution to the total capacitance;[21, 22, 23] when the electrode's QC is much larger than the electrolytic capacitance, the total capacitance will be predominated by the ionic contribution[16, 23]. Thick dense stacks of 2D MOF sheets, even those that deliver large pores are not low-dimensional. Unless concentration of charge carriers in them is as low as in wide-band-gap semiconductors, they are expected to perform like bulk porous metallic electrodes; at least for the conductive MOF that we study here,[10, 24] it seems appropriate to exclude the QC contribution.

The capacitance is usually presented per unit (i) surface area (*area-specific*), (ii) mass (*gravimetric*), or (iii) volume (*volumetric*) of the electrode. The gravimetric and volumetric values are easy to define, whereas the determination of the area-specific capacitance could be ambiguous, as it depends how the interior surface of the electrode was measured. If it is done through adsorption isotherms, the result may depend on approximation for the isotherm, and the size of the molecular probe – with smaller probes one might report higher surface areas than that could be accessible to ions. The way how we estimated 'surface area' of the pores of the studied MOFs is described in Methods.

Fig. 4a shows the area-specific differential capacitance. The MOF with the smallest pore diameter (0.81 nm) displays a camel-like shape of the capacitance-potential curve with two maxima of 10.2 and 8.8 $\mu F\ cm^{-2}$ at -1.1 and +1.5 V, respectively, while the curves for the other two MOFs are both bell-shaped with a maximum near PZC. Within a potential range of -0.5 to +0.5 V, the MOF with pore diameter of 1.57 nm delivers the capacitance of ~10 $\mu F\ cm^{-2}$, which is compatible with RTIL-based porous carbon EDLCs.[17, 25]

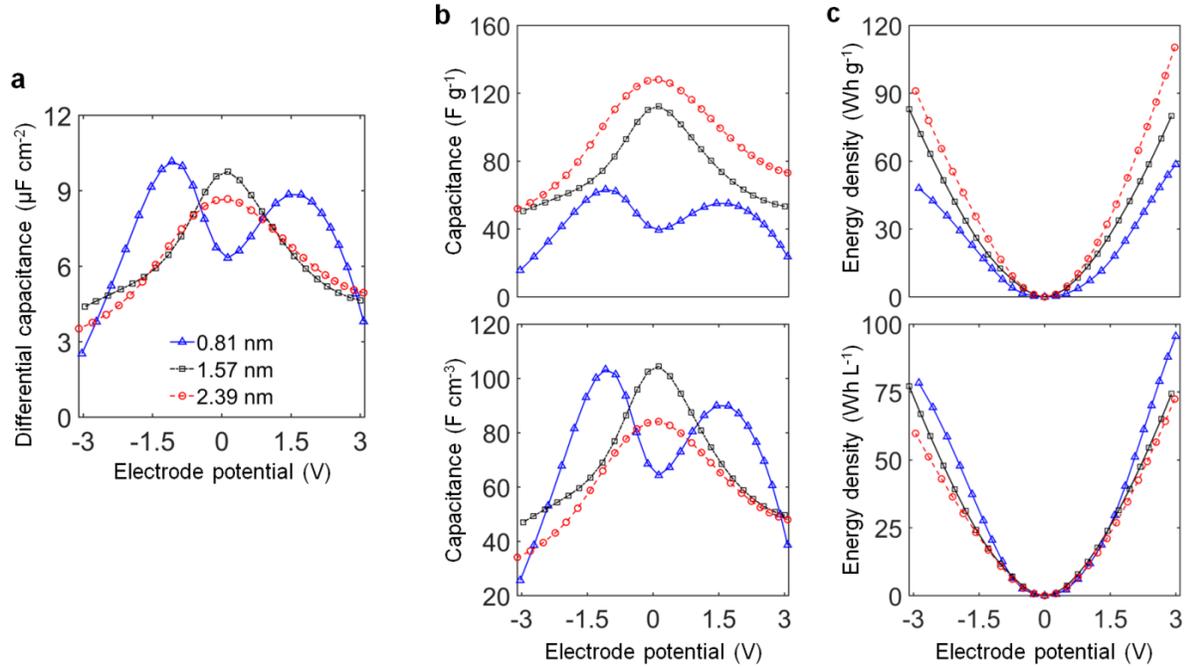

**Fig. 4. Capacitance and energy density.** Voltage dependence of the capacitance of ionic-liquid-filled MOF electrode and the stored energy density. **a,** Differential capacitance per unit pore surface area. **b-c,** Gravimetric (**b**) and volumetric (**c**) capacitance and energy density of three studied MOFs. The energy density is calculated by $E_{g/v}(\varphi) = \int \varphi C_{g/v}(\varphi) d\varphi$, where $C_{g/v}$ is the gravimetric/volumetric capacitance.

Having defined the mass and volume of a unit cell for the given structure of MOF, we obtain the gravimetric and volumetric capacitance, and the corresponding energy densities that can be stored in our MOF-based EDLCs. As shown in Fig. 4b-c, the MOF with the largest pore diameter (2.39 nm) has the highest gravimetric capacitance but lowest volumetric one. For $Ni_3(HITP)_2$ with 1.57 nm pore, the gravimetric and volumetric capacitances reach 112 F g$^{-1}$ and 105 F cm$^{-3}$, respectively, which are close to those experimentally measured for an EDLC made of this MOF and an organic electrolyte of tetraethylammonium tetrafluoroborate ([TEA][BF$_4$]) in acetonitrile (ACN).[10] Noteworthy, in our simulation of MOFs-in-RTIL system we show that, in principle, a gravimetric energy density of ~57 Wh kg$^{-1}$ could be achieved at a cell voltage of 4 V, if the electrodes could sustain this voltage. Such energy density is as high as some values reported for high-energy-density carbon EDLCs.[26] With regards to volumetric energy density, especially within a small potential range, three MOF electrodes display similar capacitances, while at 4 V working voltage, the MOF with the smallest pore size can expectedly store more energy, showing the highest volumetric energy density among these three MOFs, ~50 Wh L$^{-1}$. This predicted volumetric performance demonstrates promising potential of these MOFs in comparison with other electrode materials in EDLCs.[27, 28] Had [EMIM][BF$_4$] sustained for the given MOF a higher working voltage (5.3 V)[29], the corresponding gravimetric and volumetric energy densities could reach about 80 Wh kg$^{-1}$ and 75 Wh L$^{-1}$, respectively. Accordingly, RTILs with even wider electrochemical window (e.g., 5.9 V for 1-butyl-1-methylpyrrolidinium bis(trifluoromethylsulfonyl)imide ([pyr14][TFSI])[30] and 6.5 V for 1-butyl-1-methylazepanium bis(trifluoromethylsulfonyl)imide ([C$_4$mazp][TFSI])[31]) could be expected to further increase the energy density of EDLCs with conductive MOF electrodes that could withstand such working voltages.

The shape of the capacitance-potential curves could be understood via systematic analysis of the voltage-dependent ion distributions inside the pores – cf. Figs 2 and 3 as well as Supplementary Figure 7. There, for the smallest pore MOF (0.81 nm), the number of in-pore cations and anions, separately, changes slowly within the potential range between -1 and +1.4 V. The slope gets steeper under larger electrode polarizations, slowing down after ±2 V, approaching saturation. This behavior results in the camel-shape of capacitance-potential curve (Fig. 4a). For the other two MOFs (Supplementary Figure 7b-c), the steepest change in ion population occurs near PZC, slowing down as electrodes become more polarized (especially, over ±1.5 V); this gives rise to bell-shaped capacitance-potential curves (Fig. 4a). Notably, for pores filled with more than one layer of ions, within the potential range of -0.5 to +0.5 V the change of cation/anion population takes place majorly in the central region of a pore, but it shifts towards the pore surface region beyond this range (Supplementary Figure 7d-e).

**Charging dynamics**

We then focus on the charging dynamics – the key issue for the power delivery. Figure 5a shows the time evolution of ionic charge in a pore at 400 K (results for temperatures from 300 to 400 K can be found in Supplementary Figure 8). It appears possible to rationalize the charging dynamics through the **t**ransmission **l**ine **m**odel (TLM).[32] Based on TLM schematized in Supplementary Figure 9a, the time evolution of the net charge of the pore surface, after applying a constant potential, reads:[32]

$$Q(t) = Q_\infty \left\{ 1 - \frac{2}{\pi^2} \sum_{n=0}^{\infty} \frac{\exp\left[-\pi^2 \left(n+\frac{1}{2}\right)^2 \left(\frac{2l}{L}\right)^2 \frac{t}{\tau}\right]}{\left(n+\frac{1}{2}\right)^2} \right\} \quad (1)$$

where $Q_\infty$ is the value of the net surface charge when the pore gets fully charged, $l$ is the pore volume divided by its surface area, and $L$ is the full length of the pore. The parameter in this equation, which does not depend on pore length, is the intrinsic relaxation time,

$$\tau = \frac{C_{area} \times l}{\sigma} \tag{2}$$

in which $C_{area}$ is the capacitance per unit surface area of the pore and $\sigma$ is the ionic conductivity inside the pore. Note that Eq. (1) is strictly valid for potential-independent capacitance; there is no closed-form solution if the capacitance varies during charging. For simplicity, we will still use Eqs. (1) and (2) with the value of $C_{area}$ corresponding to the integral capacitance for a given electrode potential.

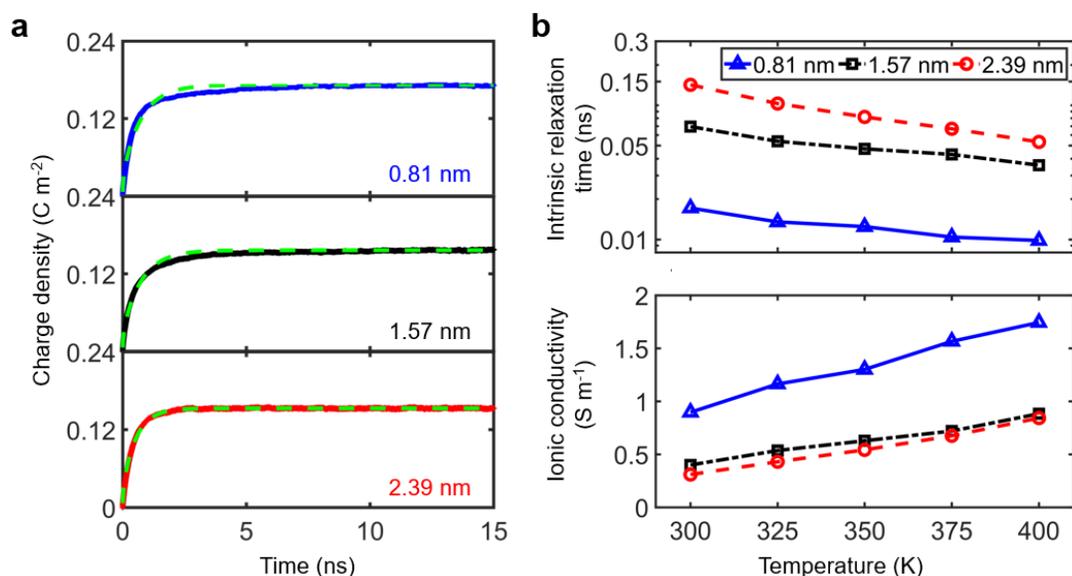

**Fig. 5. Charging process at nanoscale. a**, Time evolution of charge density per unit surface area of the pore, after a cell voltage of 4 V was applied between two electrodes, shown for the positive electrode, for three MOFs of indicated pore sizes at 400 K. MD-obtained results are shown against the curves (green dashed lines) fitted by the transmission line model sketched in Supplementary Figure 9a. **b**, Intrinsic relaxation time (top panel, shown in logarithmic scale for better visibility) and the ionic conductivity (bottom panel) of RTILs in the MOF pores at different temperature.

Taking 1.57 nm pore-diameter MOF, with $l = 0.40$ nm and $L = 5.86$ nm (sets of parameters for all three studied MOFs can be seen in Supplementary Tables 2 and 3), by fitting the TLM to simulation data we obtain the intrinsic relaxation time, $\tau$. Regardless of the aforementioned approximation in Eq. (1), TLM-fitted curves match the MD-obtained charging dynamics very well (see Fig. 5a). This is further confirmed by fittings of the charging dynamics under alternative working voltages and at different temperatures (Supplementary Figures 8 and 9b).

Doing such fittings for all three studied MOFs shows that with increasing temperature, $\tau$

decreases (top panel of Fig. 5b), which is reasonable and has been experimentally demonstrated for porous electrodes owing to the increase of ion mobility with temperatures[33, 34]. The obtained value of $\tau$ can be used to estimate the charging time, $\tau_0 = \frac{\tau}{(l/L)^2}$, for a practical supercapacitor cell. For instance, for 1.57 nm MOF electrode of 100 μm thickness at 300 and 400 K, $\tau_0$ will be 4.3 and 2.2 seconds, respectively.

Performing MD simulations for **e**lectrical **c**urrent **a**uto-**c**orrelation **f**unctions (ECACFs) and using Green-Kubo formula for electrical conductivity (see Methods), we obtained for unconfined [EMIM][BF$_4$] the bulk conductivity of 1.26 – 8.2 S m$^{-1}$ with the temperature varying from 300 to 400 K, consistent with experimental data (1.6 – 10.8 S m$^{-1}$).[35] But how and to what extent would the temperature affect the conductivity of RTIL inside MOF pores? We obtain the values of $\sigma$ in the following way: we extract the values of $\tau$ from fitting Eq. (1) to the MD-obtained charging curves in Fig. 5a, and then evaluate $\sigma$, using Eq. (2). For the MOFs with the pore size of 1.57 nm and 2.39 nm, the obtained values of $\sigma$ increase from 0.3 – 0.9 S m$^{-1}$ within the temperature range of 300 – 400 K (bottom panel of Fig. 5b). Interestingly, the conductivity of ions in the smallest MOF pore (0.81 nm), varying within 0.9 – 1.7 S m$^{-1}$ in such temperature interval, appears higher than in the other two larger MOF pores. This could possibly be attributed to the superionic state of the RTIL confined in the smallest pore, in which electrostatic interactions between ions are partially screened out due to the induced image charges on the pore wall.[36, 37] Nevertheless, the conductivities of ions under nanoconfinements are obviously smaller than in the bulk; their temperature dependence is less pronounced: inside MOF it is about 2 – 3 times, against 6.5 times increase in the bulk, and it becomes weaker for the narrower-pore MOFs.

These results suggest considerable limitations for ion transport inside the quasi-1D nanopores of MOFs. Hence, the electrolyte resistance in MOFs would be the dominant contribution to the equivalent series resistance (ESR) of a practical MOF-based EDLC cell. This is similar to what has been concluded from the experiment with 1.57 nm pore-diameter MOF in an organic solution.[10] It is simply because that the electrical conductivity of this MOF is about three orders of magnitude higher than the ionic conductivity of RTIL. Interestingly, the larger working voltage results in a decrease of $\tau$ and the increase of $\sigma$ (cf. Supplementary Figure 9e-g), thus promoting faster ion transport inside MOFs.

**Capacitive performance at macroscale**

Thus obtained parameters, the capacitance and the conductivity of electrolyte in a MOF pore, together with some intrinsic properties of MOFs (i.e., their specific surface area, density and porosity, cf. Supplementary Table 2), could be used to assess the capacitive performance of a 'practical' MOF-based supercapacitor, via macroscale equivalent circuit simulations (for details see Methods and Supplementary Part 5 with Supplementary Figure 10). Based on the values of $\sigma$ and capacitance, Nyquist plots in Fig. 6a were computed for the supercapacitors with 100 μm thick electrodes,13 mm diameter of electrodes, and 25 μm separator, as benchmarks for impedance spectroscopy. From there, we obtain ESRs of such cells as 1.35, 1.22 and 0.98 Ω for MOFs with the pore diameter of 0.81, 1.57 and 2.39 nm, respectively, at 400 K. At 300 K, ESR increases to approximately 2.7 Ω for all three MOF electrodes, which is much smaller than that (8.6 Ω) of carbon-based supercapacitor with RTIL [EMIM][TFSI].[38] It was earlier experimentally demonstrated that the electrodes based on aligned single-walled carbon nanotubes show greatly enhanced ion transport parallel to the alignment direction.[39] Similarly, the MOFs studied herein, possessing crystal structure and retaining crystallinity when made into electrodes,[10, 24] could offer straight quasi-cylindrical pores, providing fast charging dynamics.

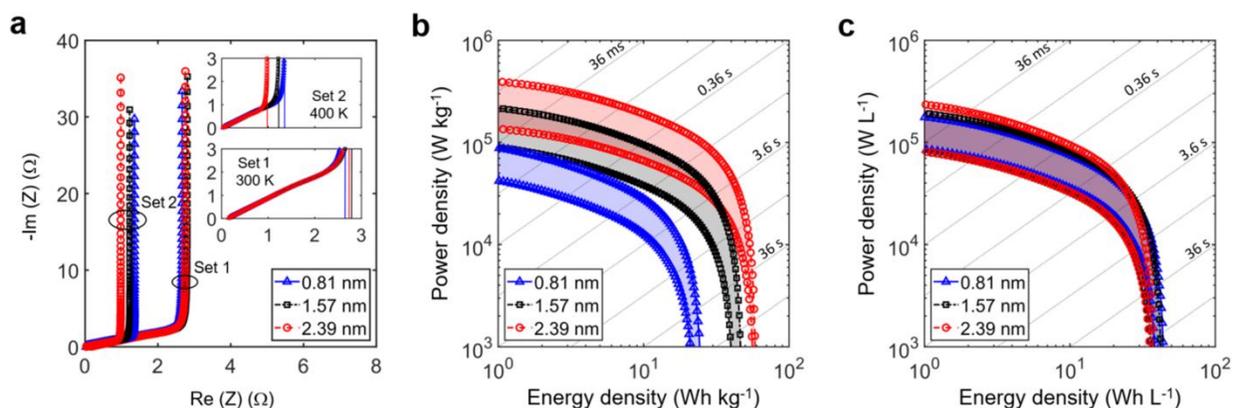

**Fig. 6. Capacitive performance of practical cell-size supercapacitors**. **a,** Nyquist plots for MOF-based supercapacitors at temperatures of 300 K (Set 1) and 400 K (Set 2). **b-c,** Gravimetric (**b**) and volumetric (**c**) Ragone plots for MOF-based supercapacitors from 300 to 400 K. Blue triangles, black squares, and red dots represent results for MOFs with the pore sizes of 0.81, 1.57, and 2.39 nm, respectively. Grey dotted lines in **b-c** indicate the lasting time that quantifies how long a supercapacitor can supply the power at an appointed power-energy point. The voltage between cathode and anode is 4 V.

The confluence of charging dynamics and charge storage of practical MOF-based supercapacitors is demonstrated by Ragone plots for temperature-dependent power-energy relationships shown in colored regions in Fig. 6b-c. Specifically, the temperature has minor influence on maximal energy density but a large effect on the power density. For applications demanding the best gravimetric performance, MOFs with the largest pore size (2.39 nm) seem to be a better choice, delivering both higher energy and power density, in comparison with the other two MOFs (Fig. 6b). As the temperature increases from 300 to 400 K, under a cell voltage of 4 V, the MOF with the largest pore size could reach power density of 135 – 390 kW kg$^{-1}$ and an energy density about 57 Wh kg$^{-1}$. For optimized volumetric performance, all three MOFs exhibit similar promising performance, with a power density range of 75 – 215 kW L$^{-1}$ and energy density of ~40 Wh L$^{-1}$. These performances compare favorably with most reported carbon-based EDLCs (for detailed comparison see Supplementary Table 4).[26, 27, 28, 40, 41] As expected, under smaller voltages, the energy and power densities would decrease (cf. Supplementary Figure 11).

## Conclusion

We have investigated the charge storage and charging dynamics of supercapacitors consisting of conductive MOF electrodes and RTIL electrolyte. The microstructures of RTIL inside MOF pores were analyzed in terms of the in-plane and axial ion distributions, as well as ion orientations, which helps to interpret the obtained camel- and bell-shapes of the capacitance-potential dependence. The charging dynamics has been rationalized within the transmission line model, which we further used to evaluate the capacitive performance of a sample EDLC device at the macroscale and with the temperature dependence, based on the MD-obtained information. Modeling results revealed that these MOF/RTIL-based cells could exhibit capacitive performance superior to most carbon-based devices,[26, 27, 28, 40, 41] which suggest promising avenues for designing supercapacitors *with both high energy and power densities, especially when volumetric performance is concerned*.

To be developed 3D conductive MOFs scaffolds might be expected to have advantages over the dense stacks of 2D-MOF-sheets with quasi-1D pores. Indeed, 3D scaffolds could provide ion transport paths in all directions and consequently promote cation-anion swapping, pivotal in charging-discharging processes. Such scaffold-electrodes with the all-dimension openness and high porosity may help to enhance the charging dynamics, and together with the enhanced surface area, potentially enlarge energy and power densities simultaneously.

## Methods

### Molecular dynamics simulation

As shown in Fig. 1, the MD simulation system consists of two identical and symmetric conductive MOFs immersed in a RTIL [EMIM][BF$_4$]. Real supercapacitors certainly contain an electronically isolating ion transport membrane that warrants the absence of a short-cut between the electrodes, however, ideal membranes must not impede ion exchange between the electrodes. For this proof-of-the principle study, considering therefore the best performance possible, we will not incorporate the membrane into our simulation cell. This strategy is generally adopted for MD modeling of supercapacitors.[15, 17, 32, 42, 43] The atomistic structures of MOFs were obtained from experimental measurements.[24] The geometry optimization of each MOF was derived from density function theory (DFT) calculations, using in Vienna ab initio simulation package (VASP)[44]. Details of MOF structure optimization of and pore size calculation can be seen in Supplementary Part 1 and Supplementary Tables 1-2. The Lennard-Jones parameters for the MOF atoms were taken from the generic universal force field (UFF)[45]; the coarse-grained model was adopted for [EMIM][BF$_4$], which could provide the proper thermodynamic and dynamic properties[42]. The simulation cells were chosen as large enough to reproduce the bulk state in the central region of RTIL reservoir connected with two electrodes, and periodic boundary conditions were applied in all directions. Specific system parameters are given in Supplementary Table 3.

Simulations were performed in the NVT ensemble using a customized MD software GROMACS[46]. The applied electrical potential between the two electrodes in Fig. 1 was maintained by the constant potential method (CPM), as it allows the fluctuations of charges on electrode atoms during the simulation.[15, 42, 43] Details of CPM could be found in Supplementary Part 2. To guarantee the accuracy, the electrode charges are updated on the fly of simulation running at every simulation step (2 fs). The electrolyte temperature was maintained at 400 K using the V-rescale thermostat[47]. The electrostatic interactions were computed using the particle mesh Ewald method[48]. An FFT grid spacing of 0.1 nm and cubic interpolation for charge distribution were used to compute the electrostatic interactions in the reciprocal space. A cutoff length of 1.2 nm was used in the direct summation of the non-electrostatic interactions and electrostatic interactions in the real space. For each conductive MOF, the MD system was annealed from 500 to 400 K over a period of 10 ns, following by running another 40 ns to reach equilibrium under null electrode potential. To explore the charging dynamics, five independent runs were performed for smoothing the charging process data. To obtain microstructure and capacitance, a simulation was performed for 60 ns to surely reach equilibrium under the applied potential ranging from 0 to 6 V, and then another 60 ns production in equilibrium state was run for analysis.

The electrical conductivity of the bulk [EMIM][BF$_4$] was evaluated via the time integral of ECACFs, $\sigma_{RTIL} = \frac{1}{3Vk_BT}\int_0^\infty <\vec{J}(0)\cdot\vec{J}(t)>dt$, in which $V$ is the system volume and $\vec{J}(t)$ denotes the electrical current. $\vec{J}(t)$ is defined by $\vec{J} = \sum_{i=1}^{N} q_i\vec{v}_i$, where $N$ is the total number of ions, and $q_i$ and $\vec{v}_i(t)$ are the charge and velocity of the $i$-th ion, respectively.

It is worth noting that static prosperities of EDLCs were studied at a temperature of 400 K that is generally used for MD simulations to get the ion structure and capacitance[32, 42], while the charging dynamics was ascertained within a temperature range of 300 – 400 K due to the big impact of temperature on dynamic properties[33, 34].

**Equivalent electrical circuit simulation**

We conceived a practical size two-electrode symmetrical cell (see Supplementary Figure 10a), based on an experimental EDLC cell with Ni$_3$(HITP)$_2$ electrode[10]. Specifically, the diameter of the MOF electrode was set as 13 mm, and the electrode thickness was taken as 100 μm for all three studied MOFs. The equivalent circuit model of the conceived two-electrode symmetrical cell was based on RC transmission line circuit in which the resistors and the capacitors were calculated based on the size of the cell and the MD-obtained specific capacitance and conductivity of ions in MOF. The circuit simulations were performed via *Simulink*, in which impedance measurement and constant power load (via boosting DC-DC converter block) tests were carried out respectively to obtain the Nyquist plots and Ragone plots for all three MOF-based EDLCs in practical cell size. Detail can be seen in Supplementary Part 5 with Supplementary Figure 10.